\documentclass[pdftex,twocolumn,epjc3]{svjour3}          
\RequirePackage[T1]{fontenc}

\smartqed  

\RequirePackage{graphicx}
\RequirePackage{mathptmx}      
\RequirePackage{flushend}
\RequirePackage[numbers,sort&compress]{natbib}
\RequirePackage[colorlinks,citecolor=blue,urlcolor=blue,linkcolor=blue]{hyperref}

\graphicspath{{figures/}}
\RequirePackage{subfigure}
\RequirePackage{mathtools}
\RequirePackage{amsmath}
\RequirePackage{lineno}
\RequirePackage{textgreek}
\RequirePackage{multirow}
\allowdisplaybreaks

\journalname{Eur. Phys. J. C}

\begin{document}
\title{Performance of the SABRE detector module in a purely passive shielding}

%

\author{F.~Calaprice\thanksref{princeton} \and 
J. B.~Benziger\thanksref{princeton} \and 
S.~Copello\thanksref{genova} \and 
I.~Dafinei\thanksref{roma1infn} \and 
D.~D'Angelo\thanksref{milanoinfn,milano} \and 
G.~D'Imperio\thanksref{roma1infn} \and 
G.~Di~Carlo\thanksref{lngs} \and 
M.~Diemoz\thanksref{roma1infn} \and 
A.~Di Giacinto\thanksref{lngs} \and 
A.~Di~Ludovico\thanksref{lngs} \and 
M.~Ianna\thanksref{milano} \and 
A.~Ianni\thanksref{lngs} \and 
A.~Mariani\thanksref{e1,princeton,lngs} \and 
S.~Milana\thanksref{roma1infn} \and 
D.~Orlandi\thanksref{lngs} \and 
V.~Pettinacci\thanksref{roma1infn} \and 
L.~Pietrofaccia\thanksref{lngs} \and 
S.~Rahatlou\thanksref{roma1infn,roma1} \and 
B.~Suerfu\thanksref{berkeley} \and 
C.~Tomei\thanksref{roma1infn} \and 
C.~Vignoli\thanksref{lngs} \and 
A.~Zani\thanksref{milanoinfn}
}

\thankstext{e1}{e-mail: am8894@princeton.edu (corresponding author)}

\institute{Physics Department, Princeton University, Princeton, NJ 08544, USA\label{princeton} \and
Dipartimento di Fisica, Universit{\`a} degli Studi di Genova and INFN Genova, Genova I-16146, Italy\label{genova} \and
INFN - Sezione di Roma, Roma I-00185, Italy\label{roma1infn} \and
INFN - Sezione di Milano, Milano I-20133, Italy\label{milanoinfn} \and
Dipartimento di Fisica, Universit{\`a} degli Studi di Milano, Milano I-20133, Italy\label{milano} \and
INFN - Laboratori Nazionali del Gran Sasso, Assergi (L'Aquila) I-67100, Italy\label{lngs} \and
Dipartimento di Fisica, Sapienza Universit{\`a} di Roma, Roma I-00185, Italy\label{roma1} \and
University of California Berkeley, Department of Physics, Berkeley, CA 94720, USA\label{berkeley}
}

%
%
%
\date{Received: date / Revised version: date}
%

\maketitle

\begin{abstract}

We present here a characterization of the low background NaI(Tl) crystal NaI-33 based on a period of almost one year of data taking (891 kg$\times$days exposure) in a detector configuration with no use of organic scintillator veto. 
This remarkably radio-pure crystal already showed a low background in the SABRE Proof-of-Principle (PoP) detector, in the low energy region of interest (1-6 keV) for the search of dark matter interaction via the annual modulation signature. 
As the vetoable background components, such as $^{40}$K, are here sub-dominant, we reassembled the PoP setup with a fully passive shielding. 
We upgraded the selection of events based on a Boosted Decision Tree algorithm that rejects most of the PMT-induced noise while retaining scintillation signals with $>$ 90\% efficiency in 1-6 keV. 
We find an average background of 1.39 $\pm$ 0.02 counts/day/kg/keV in the region of interest and a spectrum consistent with data previously acquired in the PoP setup, where the external veto background suppression was in place.
Our background model indicates that the dominant background component is due to decays of $^{210}$Pb, only partly residing in the crystal itself. The other location of $^{210}$Pb is the reflector foil that wraps the crystal. We now proceed to design the experimental setup for the physics phase of the SABRE North detector, based on an array of similar crystals, using a low radioactivity PTFE reflector and further improving the passive shielding strategy, in  compliance with the new safety and environmental requirements of Laboratori Nazionali del Gran Sasso.

\end{abstract}

%
%

\newcommand{\tr}[1]{\textcolor{red}{#1}}

\maketitle

\section{Introduction}

The SABRE (Sodium-iodide with Active Background REjection) experiment aims to perform an independent search for the annual modulation signature of the dark matter (DM) interaction rate with an unprecedented sensitivity to confirm or refute the DAMA claim (short for DAMA/LIBRA) \cite{dama2020-summary}. 
The ultimate goal of SABRE is to deploy two independent NaI(Tl) crystal arrays in the northern (SABRE North) and southern (SABRE South) hemispheres to identify possible contributions to the modulation from seasonal or site-related effects. The presently running ANAIS~\cite{anais-3y}
and COSINE~\cite{cosine-3y} experiments also operate NaI(Tl) crystal arrays. However they have not yet provided a conclusive answer to the longstanding debate over the DM interpretation of the DAMA  claim. This is due to their 2-3 times higher background in the 1-6 keV energy region of interest (ROI) with respect to that measured by DAMA ($\sim$ 1 count/day/kg/keV).

SABRE pursues a background rate lower than DAMA, namely of 0.3-0.5 counts/day/kg/keV (cpd/kg/keV) in the same DM ROI, and the key element in achieving this ambitious goal is the use of ultra-high purity NaI(Tl) crystal. 
In this pathway the crystal denominated NaI-33 was thoroughly characterized in 2020 inside the SABRE Proof-of-Principle (PoP) detector \cite{sabre-pop-2021} at Laboratori Nazionali del Gran Sasso (LNGS).
In addition to a passive shielding, the PoP detector was equipped with a liquid scintillator veto. This was meant to effectively suppress $^{40}$K and other gamma-emitting contaminants of the crystal contributing to the background in the ROI \cite{sabre-pop}. 
Over the last year this picture has changed.
The PoP experience has demonstrated a very low internal radioactivity and a overall background rate of 1.20 cpd/kg/keV in the ROI. 
The vetoable background components are lower than our best expectations and no longer constitute a dominant background.  
In this paper we describe how we modified the design to a vetoless configuration obtaining a comparable background level in the ROI and a shielding design fully compliant with the LNGS new safety rules. This is an enabling step toward the physics phase of the SABRE North experiment.

\section{Scintillator-free setup}

\begin{figure}[!ht]
    \centering
    \includegraphics[width=.23\textwidth]{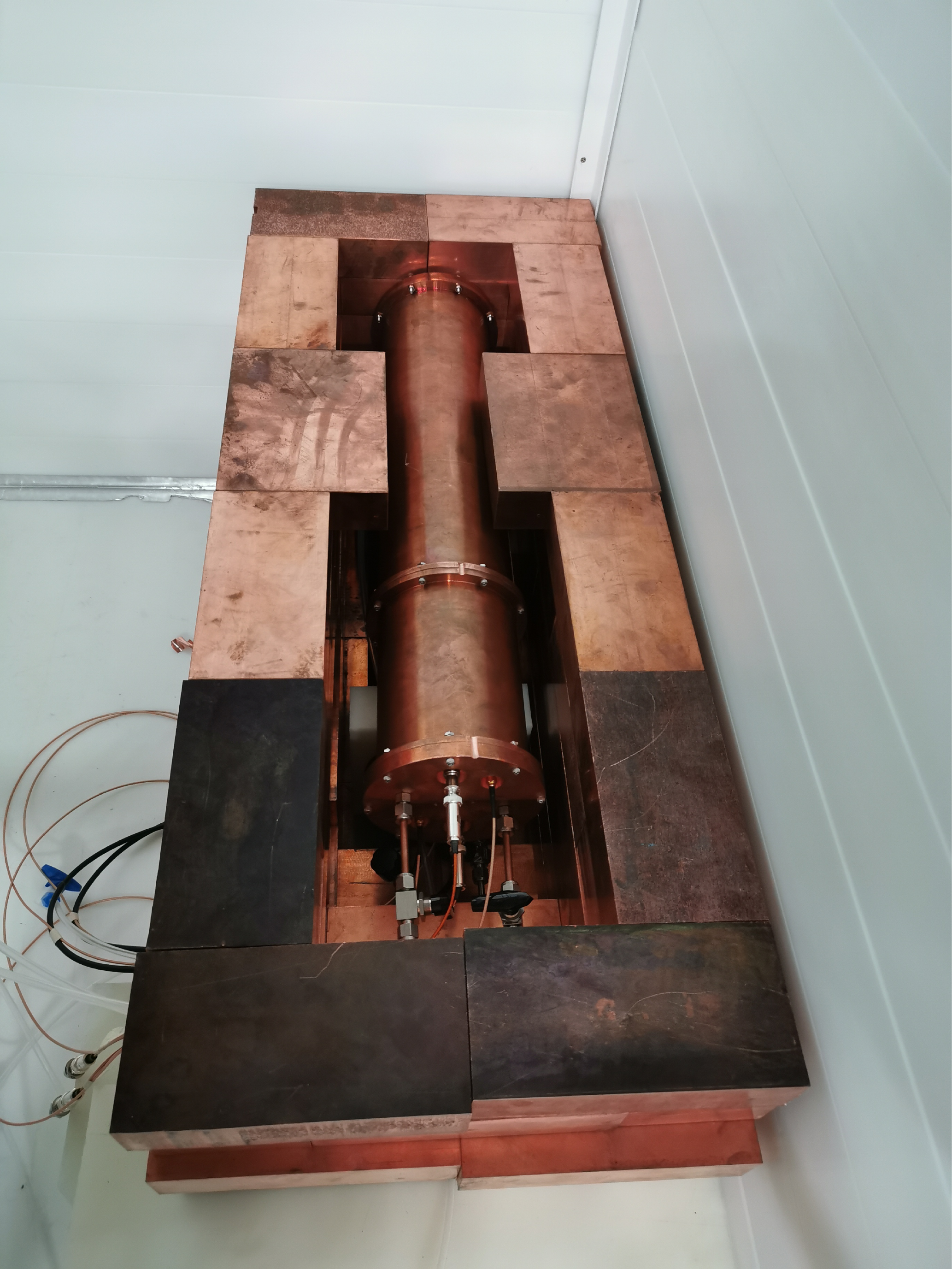}
    \includegraphics[width=.23\textwidth]{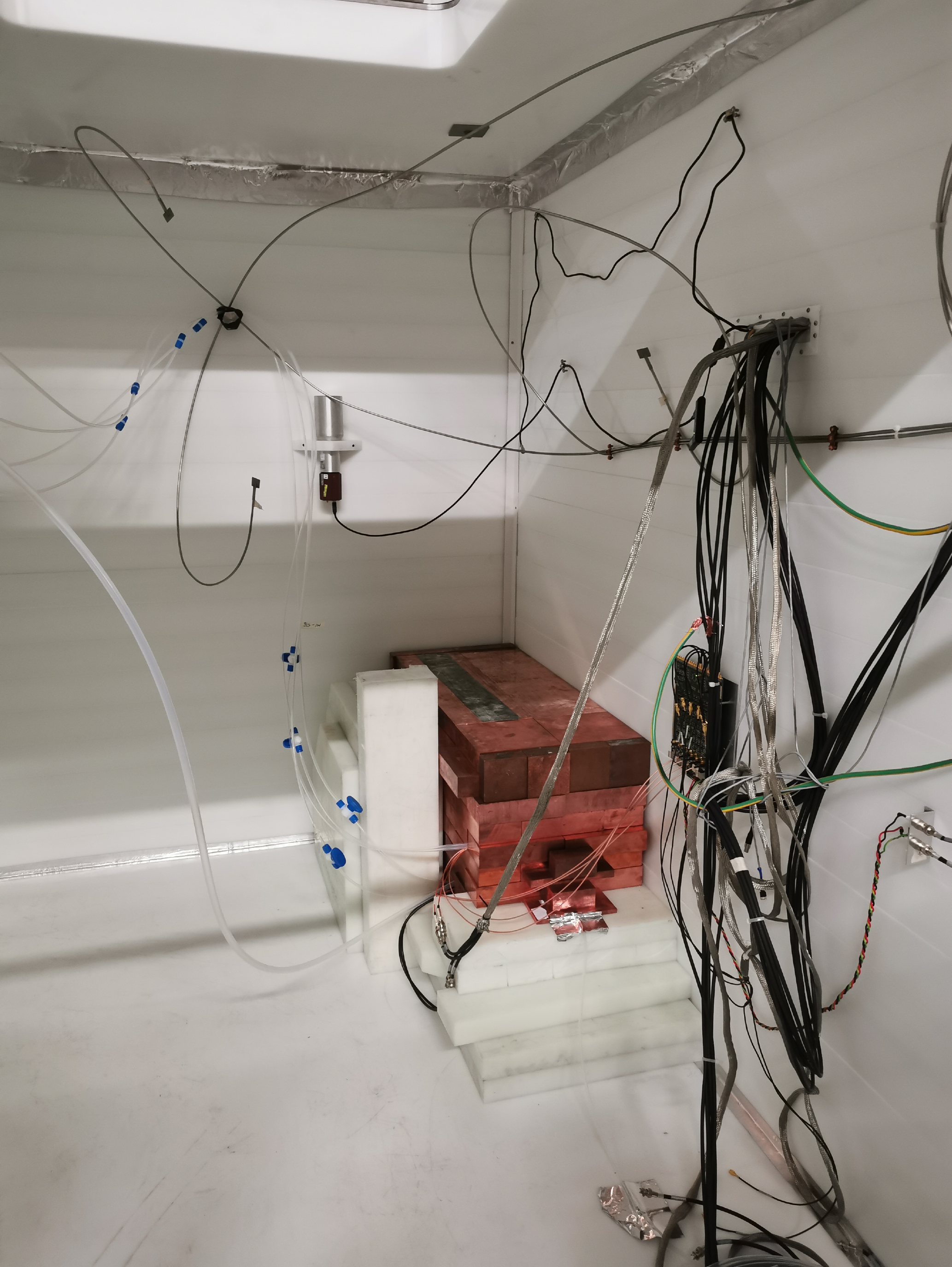}
    \caption{Left: PoP-dry copper shielding opened inside the SABRE PoP passive shielding at LNGS. The detector module containing the NaI-33 crystal is also visible in this picture. Right: PoP-dry copper + PE slabs shielding closed inside the SABRE PoP passive shielding at LNGS.}
    \label{fig:PoPdrySetup}
\end{figure}

The experimental setup described in this article stems from the SABRE-PoP~\cite{sabre-pop-2021}. 
The detector module consists of the NaI-33 crystal (3.4~kg) directly coupled to two Hamamatsu R11065-20 Photomultiplier Tubes (PMTs). 
In the PoP setup the module was placed inside a steel vessel containing a 2-ton liquid scintillator veto. 
At the beginning of 2021 we modified the PoP setup to restart the characterization of SABRE crystals without the liquid scintillator and we will be referring to it in the following as the PoP-dry. After removing the veto vessel, we placed the detector module directly inside the PoP passive shielding (polyethylene slabs plus water tanks). 
To compensate for the missing shielding power of the liquid scintillator (about 70 cm), we added a low radioactivity copper layer (10 cm on all sides and top, 15 cm below) and some additional polyethylene slabs around the copper, as can be seen in Fig.~\ref{fig:PoPdrySetup}. 
The inner volumes of the detector module and of the shielding are continuously flushed with high-purity nitrogen gas to avoid moisture and radon, respectively.

The PoP-dry Data Acquisition (DAQ) system is very similar to that used in the PoP phase. 
The data acquisition is triggered by two crystal PMT pulses exceeding 1/3 of a single photoelectron (phe) amplitude within a window of 125 ns. We have determined that the trigger inefficiency with this gate length is negligible.
Upon issuing a trigger, PMT pulses (amplified x10) are recorded for 5 $\mu$s, including a 1.5 $\mu$s pre-trigger interval and digitized at 250 MHz and 8-bit resolution with a CAEN V1720 board. 

The data analysis reported here refers to data acquired between March 17, 2021 and February 25, 2022. The duty-cycle of the data taking is $\sim$92\%, however about 16\% of the time is dedicated to calibration runs with a $^{226}$Ra source, used to tune the event selection criteria as described in the next section.

\section{Data processing and calibration}

For every PMT pulse we proceed with subtracting the baseline which is computed averaging the first 500 ns of the pre-trigger region. Next, we integrate the pulse over 1~$\mu$s starting from the trigger time. 
The number of photoelectrons is obtained dividing the charge (integral) of each pulse by the independently determined Single Electron Response of the PMT. 
The energy of the event is then obtained dividing the total number of photoelectrons of the two PMTs by the photoelectron yield of the detector module. The photoelectron yield was measured in \cite{sabre-pop-2021} and comprises the following unresolved effects: the light yield of the crystal, the light loss upon reflection on the crystal's PTFE reflector foil (side surfaces), the geometrical optical coverage and coupling, and the quantum efficiency of the PMTs. 

\begin{figure*}[!ht]
    \centering
	\includegraphics[width=.80\textwidth]{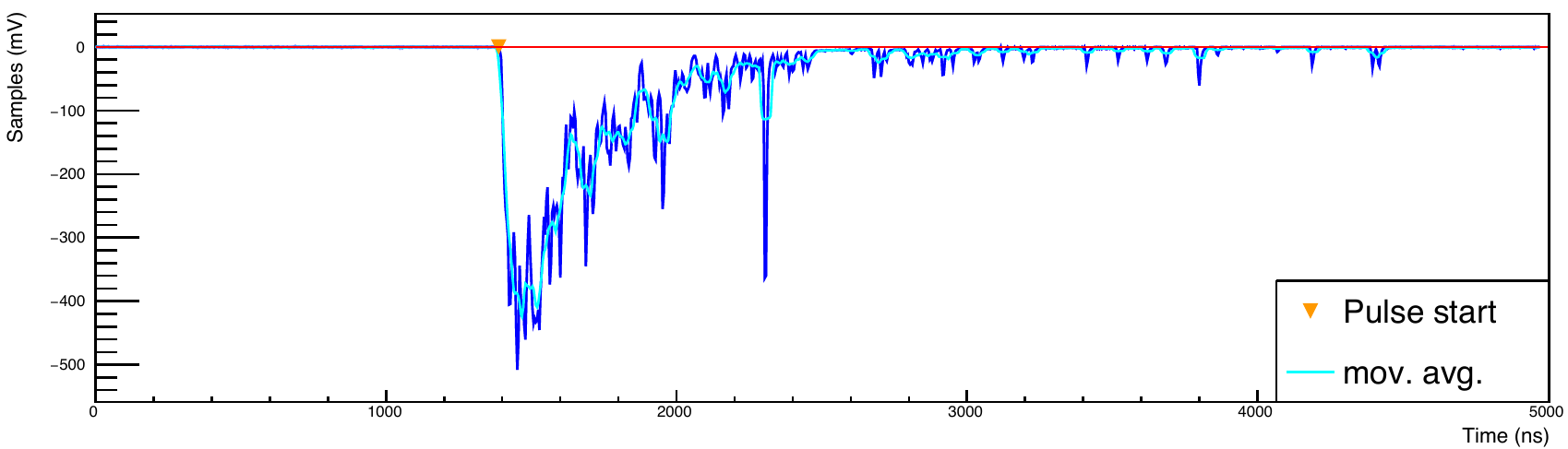}
	\includegraphics[width=.80\textwidth]{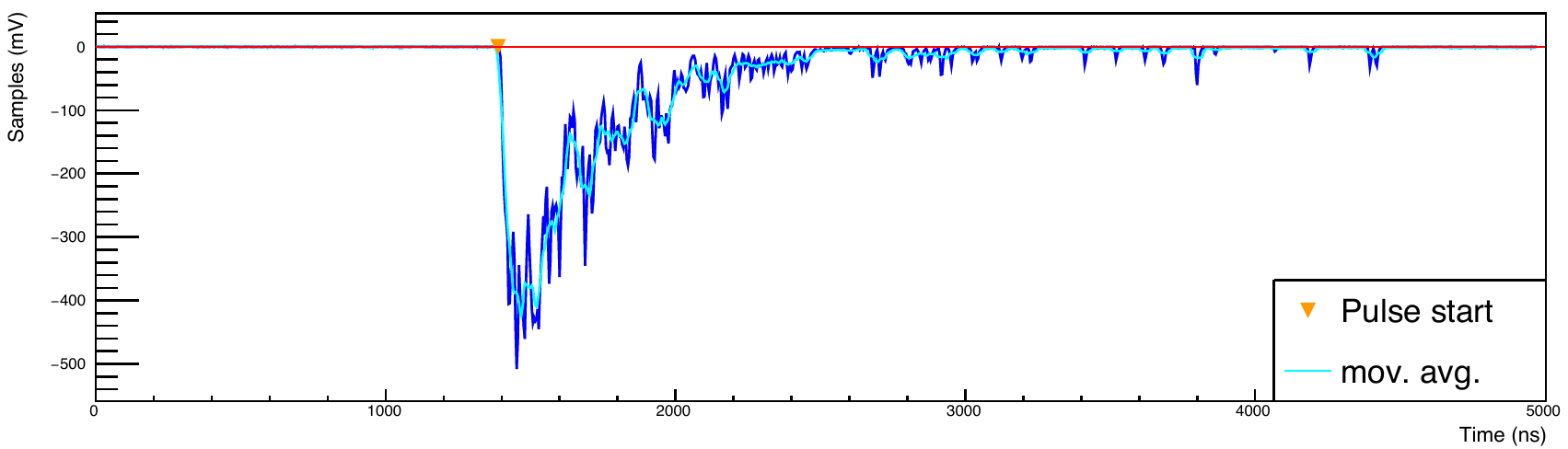}
	\caption{Example of AP suppression. (Top) Original signal acquired with the NaI-33 crystal affected by the presence of an AP, clearly visible as an anomalous spike several hundred of ns after the beginning of the main scintillation pulse. (Bottom) AP suppressed waveform.}
	\label{fig:APsuppression}
\end{figure*}

The presence of afterpulses (APs) \footnote{Fast spurious pulses that may be observed after the main signal of a PMT. One of the main sources of APs are positive ions produced by ionization of residual gases in the PMT. The positive ions return to the photocathode and produce many photoelectrons which result in APs.} may compromise the accurate energy estimation and the pulse shape discrimination. These effects are particularly important for low energy events such as those in our ROI for DM search. 
In order to address this problem, we developed an algorithm that identifies and suppresses APs \cite{master-ianna}. 
The moving average of the waveform is computed over a 100~ns time window and subtracted from the waveform itself. 
The remaining distribution of residuals is weighted for the logarithm of the moving average.
This makes the APs clearly stand out and be identified as exceeding several times the RMS of the distribution.
The AP is then suppressed by replacing the corresponding waveform samples with the average value of samples in a preceding and a following window of 40~ns each.  
Fig.~\ref{fig:APsuppression} shows an example of event that presents an AP (top) and its AP-suppressed version (bottom).

\begin{figure*}[!ht]
    \centering
	\includegraphics[width=.80\textwidth]{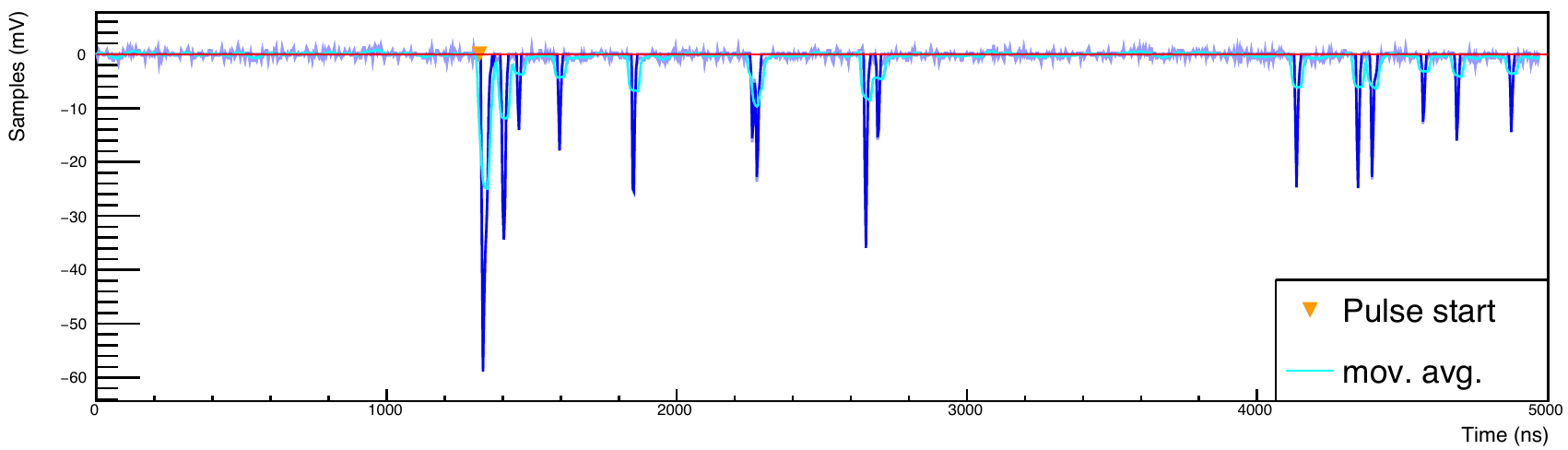}
	\includegraphics[width=.80\textwidth]{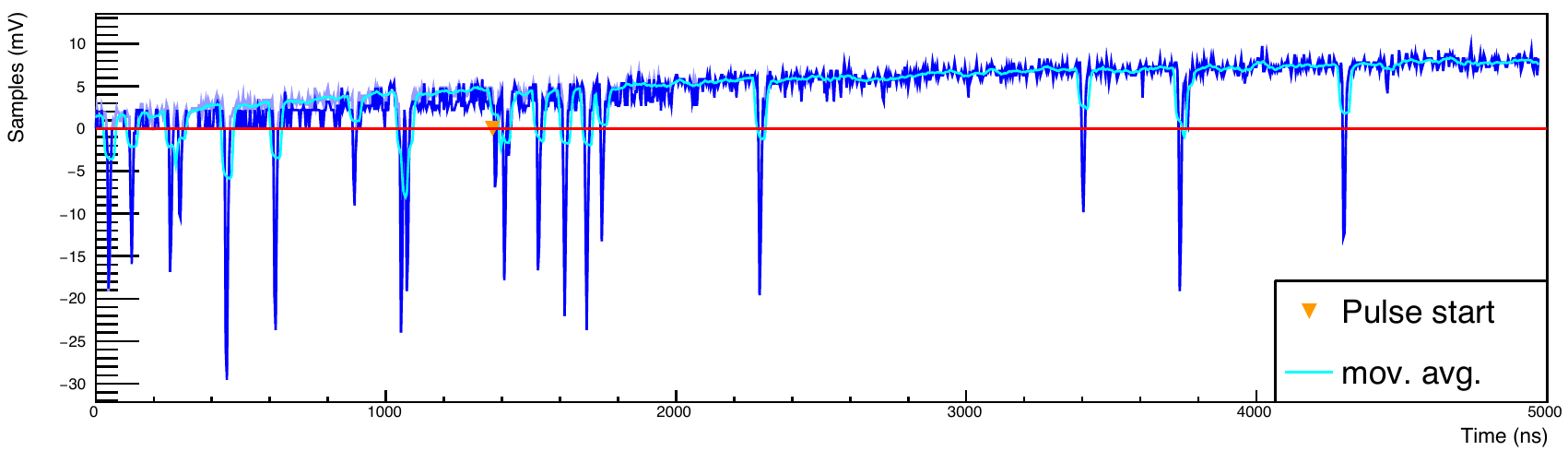}
	\includegraphics[width=.80\textwidth]{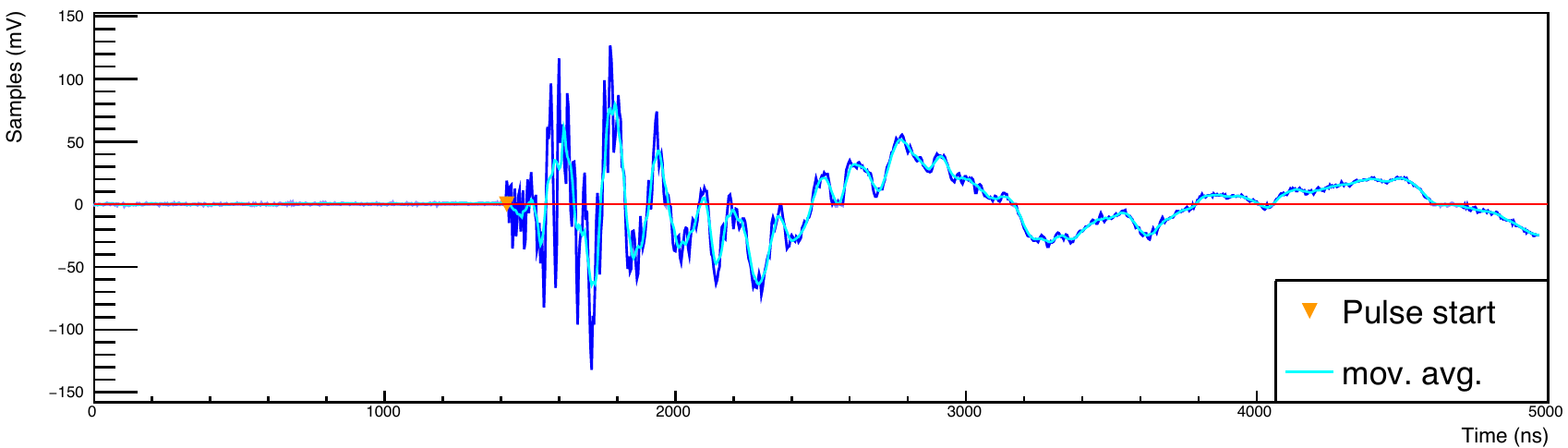}
	\caption{Examples of events rejected by the data quality cuts. (Top) Tail-event following a high energy interaction. (Center) Event with anomalous baseline estimation. The baseline is shown in red. (Bottom) Bipolar event.}
	\label{fig:EventExamples}
\end{figure*}

We apply data quality cuts to reject anomalous events that can be excluded prior the selection discussed in the next section. This preliminary operation safely preserves all signal events. 
We reject bursts of events following 
a very high energy event and that can be ascribed to the tail of its scintillation emission (Fig.~\ref{fig:EventExamples}, top). We reject events for which an accurate baseline estimation (Fig.~\ref{fig:EventExamples}, center) is not possible due to photons recorded in the pre-trigger region or noise.
Finally, we reject bipolar events (Fig.~\ref{fig:EventExamples}, bottom), caused by interference and cross-talks in the electronic chain. 
The dead time introduced by the complex of these quality cuts is computed to be below 0.01\% and therefore neglected.

\begin{figure}[!ht]
    \centering
    \includegraphics[width=.97\columnwidth]{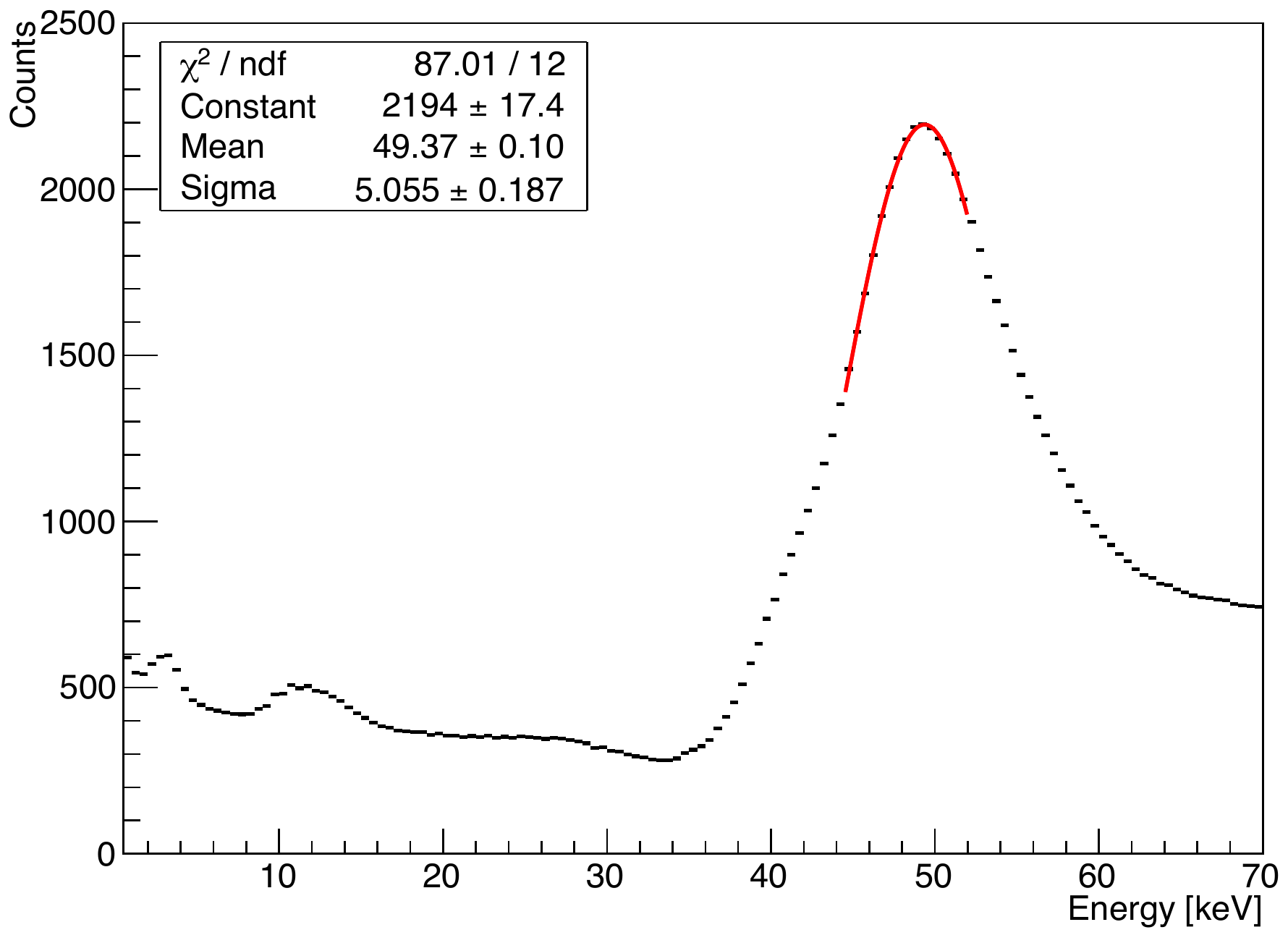}
    \caption{Gaussian fit to the $^{210}$Pb peak in the Monte Carlo simulated spectrum of crystal NaI-33. The position of this peak, calculated by the fit, is used to evaluate the energy correction factor to be applied in real data.}
    \label{fig:CalibrationPb210Peak}
\end{figure}

Calibration of the spectrum in the low energy region of interest using external sources such as $^{241}$Am may be affected by the small penetration of low energy gammas and therefore, by a misrepresentation of the full scintillating volume. 
Consequently, we exploit intrinsic $^{210}$Pb to calibrate the energy spectrum and to monitor its stability.
$^{210}$Pb $\beta$-decays to $^{210}$Bi, mostly (84\%) to an excited  state. 
The 46.5~keV de-excitation adds to the average energy of the $\beta$-decay and gives a peak-like spectral feature at $\sim$50 keV.
We assume as reference energy for the peak the mean of a Gaussian fit performed on a simulated $^{210}$Pb signal in our detector, as in Fig.~\ref{fig:CalibrationPb210Peak}. 
The GEANT4-based Monte Carlo simulation code is described in \cite{sabre-mc}, while the geometry has been modified to match the new experimental setup. 
The simulated spectrum of the energy deposited in the crystal is smeared with the measured energy resolution of the NaI-33 detector. 

In order to account for possible variations of the photoelectron yield, we measured the stability of the calibration. 
We applied a correction factor f$_{c}$ = $\mu_{\rm{data}}$/$\mu_{\rm{sim}}$, where $\mu_{\rm{data(sim)}}$ is the fitted position of the $^{210}$Pb peak in data (simulation), computed on monthly basis.
This energy correction factor ranges from 1.014 to 1.030 through our dataset, indicating photoelectron yield variations of the order of 1.6\% over a period of 11 months. 

\section{Event selection}

\begin{figure*}[!ht]
    \centering
	\includegraphics[width=.80\textwidth]{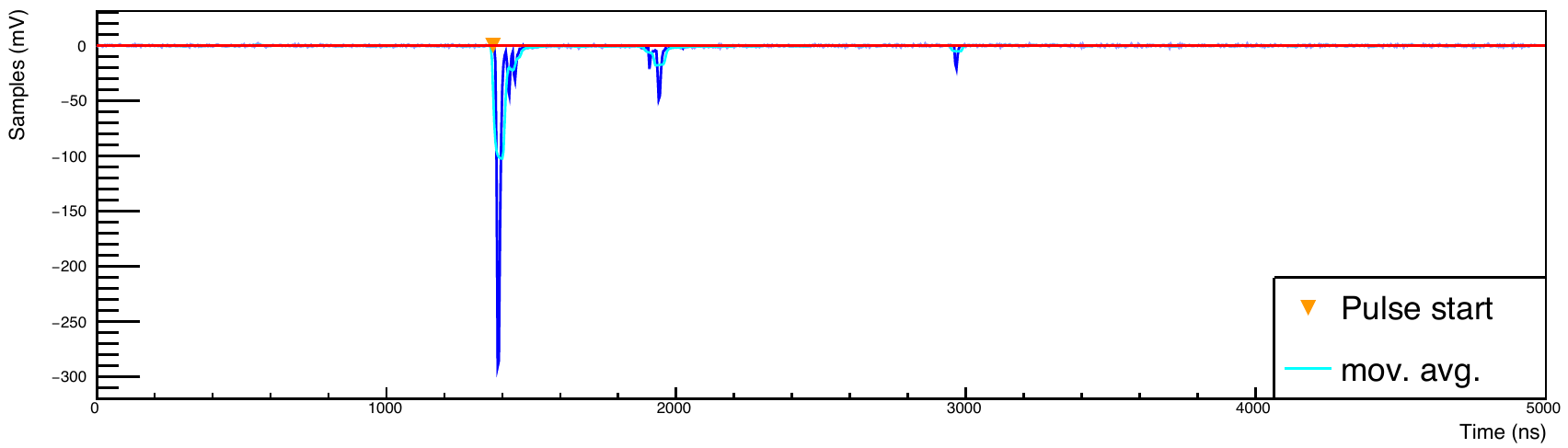}
	\includegraphics[width=.80\textwidth]{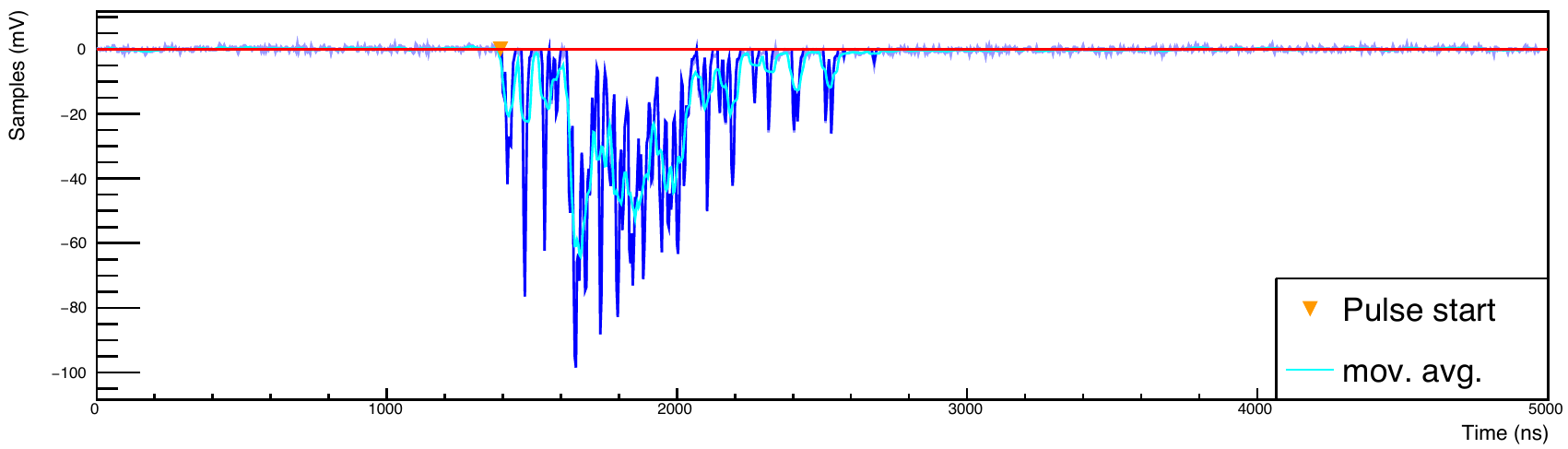}
	\includegraphics[width=.80\textwidth]{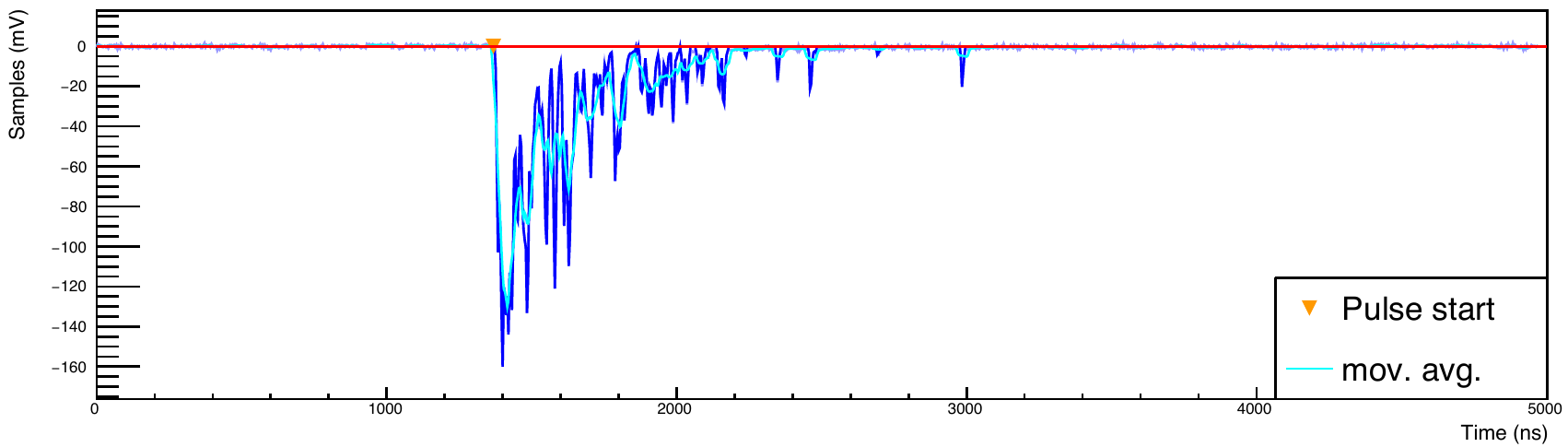}
	\caption{Pulse shape examples: $``$spike-like$''$ noise event (top); $``$bell-like$''$ noise event (middle); scintillation signal (bottom).}
	\label{fig:EventExamples2}
\end{figure*}

The process of building the energy spectrum in a NaI(Tl) detector requires the selection of events to be retained as scintillation signal and those to be discarded as noise. 
Fig.~\ref{fig:EventExamples2} shows examples of: $``$spike-like$''$ noise event (top); $``$bell-like$''$ noise event (middle); scintillation pulse (bottom).

The selection is based on a set of variables, namely: the number of clusters in each PMT, where a cluster is a set of consecutive waveform samples exceeding an amplitude threshold of $\sim$1 phe; the charge asymmetry and the trigger time delay between the signals of the two PMTs; the amplitude-weighted mean time and its Skewness and Kurtosis; the ratio between the pulse charge and its maximum amplitude; pulse shape variables such as Tail-to-total, Head-to-total, Head-to-middle, Middle-to-tail. All these variables are defined in more details in ~\cite{nai033-udg}.

In~\cite{nai033-udg} we have introduced a traditional cut-based analysis and a more innovative Boosted Decision Trees (BDT) \cite{Friedman-2001} approach that exploits the above variables to obtain high acceptance of scintillation events while efficiently rejecting the noise with a specific focus on events up to 100 keV. 
In \cite{sabre-pop-2021} we applied the simple cut-based selection, while in this work we chose to use the multivariate approach with BDT in order to maximize the signal acceptance at low energy.
Decision trees are a popular method for classification tasks in supervised machine learning. In the training process for each decision node (i.e. variable) the cut value that gives the best separation between “signal” and “background” is defined and the process is repeated until a stop criterion is met. 
Boosting is introduced to combine many trees into a stronger classifier. The trees are trained iteratively, and at each iteration events are re-weighted according to their misclassification probability. In this way, future iterations focus more on the events that the previous ones misclassified and a stronger classifier is built. 
We use the implementation of BDT with AdaBoost~\cite{FREUND1997119} in the ROOT TMVA package~\cite{ROOT-TMVA}. 
For the signal training sample we used $^{226}$Ra source data acquired by triggering on coincidences between NaI-33 and a second adjacent NaI(Tl) crystal. This provides an extremely pure sample of true scintillation events (signal). 
In order to optimize the analysis at very low energies, we trained the BDT algorithm using as signal a fraction of the source events lying in the 0.5-10~keV energy region.
In this range, we used about 30\% of the events, enough  to ensure O(100) in each 0.5-keV bin.
For the background training we simply used standard runs (w/o source). A pure enough set of noise could be achieved selecting events in the 0.5-4~keV energy region, largely dominated by noise (93\%). About 1\% of the events in this range was used for training.

\begin{figure}[!ht]
    \centering
    \includegraphics[width=.97\columnwidth]{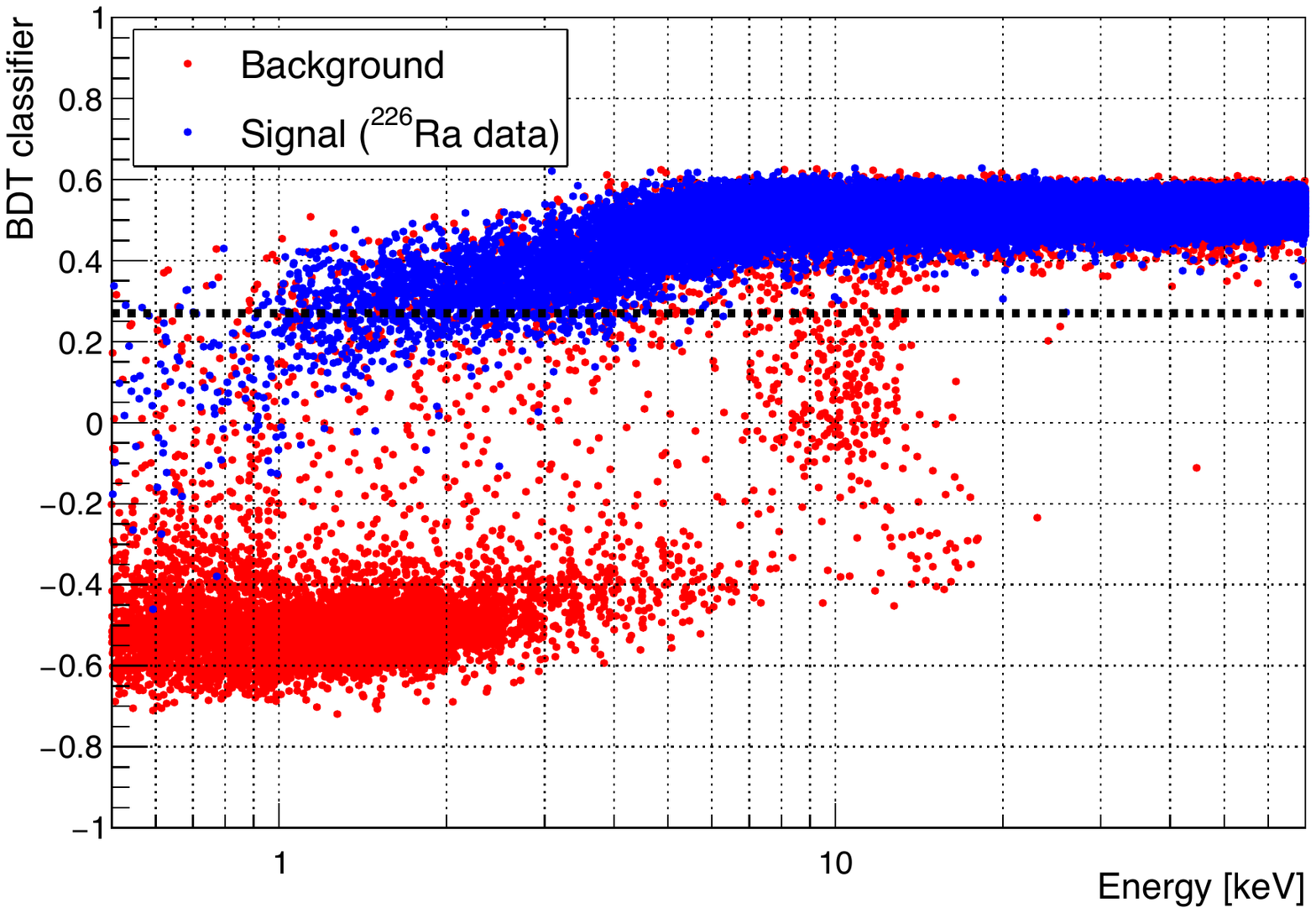}
    \caption{BDT classifier output for data acquired with the $^{226}$Ra source (blue) and background data (red), with superimposed cut threshold (black dashed line).}
    \label{fig:BDT}
\end{figure}

The BDT classifier output is defined in the range [-1,1] where lower scores are assigned to noise-like events and higher scores to signal-like events. 
Fig.~\ref{fig:BDT} shows the BDT classifier output as a function of the energy for $^{226}$Ra source data (blue) and for background data (red).  
Scintillation signals are considered those for which the BDT classifier output exceeds 0.27. This threshold was chosen based on data acquired with the $^{226}$Ra source in order to have an average event acceptance in the ROI greater than 90\%.

\begin{figure}[!ht]
    \centering
    \includegraphics[width=.97\columnwidth]{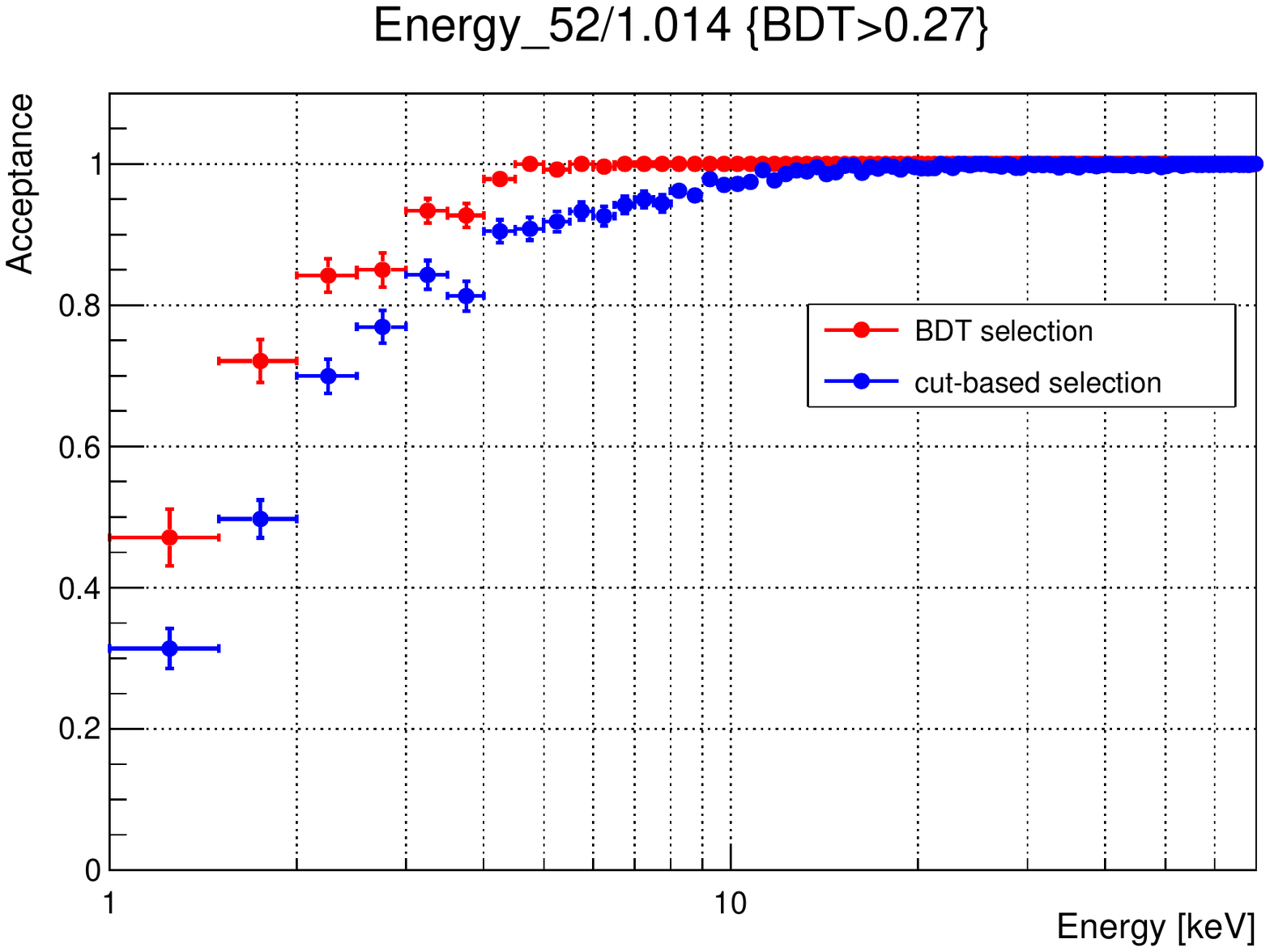}
    \caption{Acceptance as a function of the energy for the multivariate BDT analysis and, as a comparison, for the cut-based approach.}
    \label{fig:Acceptance}
\end{figure}

Fig.~\ref{fig:Acceptance} shows the signal acceptance as a function of energy below 70 keV after the BDT cut, and as comparison, also for the cut-based approach. The acceptance is measured using the $^{226}$Ra calibration data, by computing the fraction of events passing the selection. In the BDT case, events used to train the algorithm were excluded from the calculation of the acceptance to avoid bias. The multivariate analysis approach shows an acceptance that is significantly higher with respect to the cut-based method.
As a cross check, we have also performed the BDT training using a "signal" sample from a $^{176}$Lu source run. Applying this $^{176}$Lu-trained BDT with a threshold that guarantees the same 90\% acceptance, the resulting energy spectrum is compatible with the one obtained with the $^{226}$Ra-trained BDT. 

\section{Results}

\begin{figure}[!ht]
    \centering
    \includegraphics[width=.97\columnwidth]{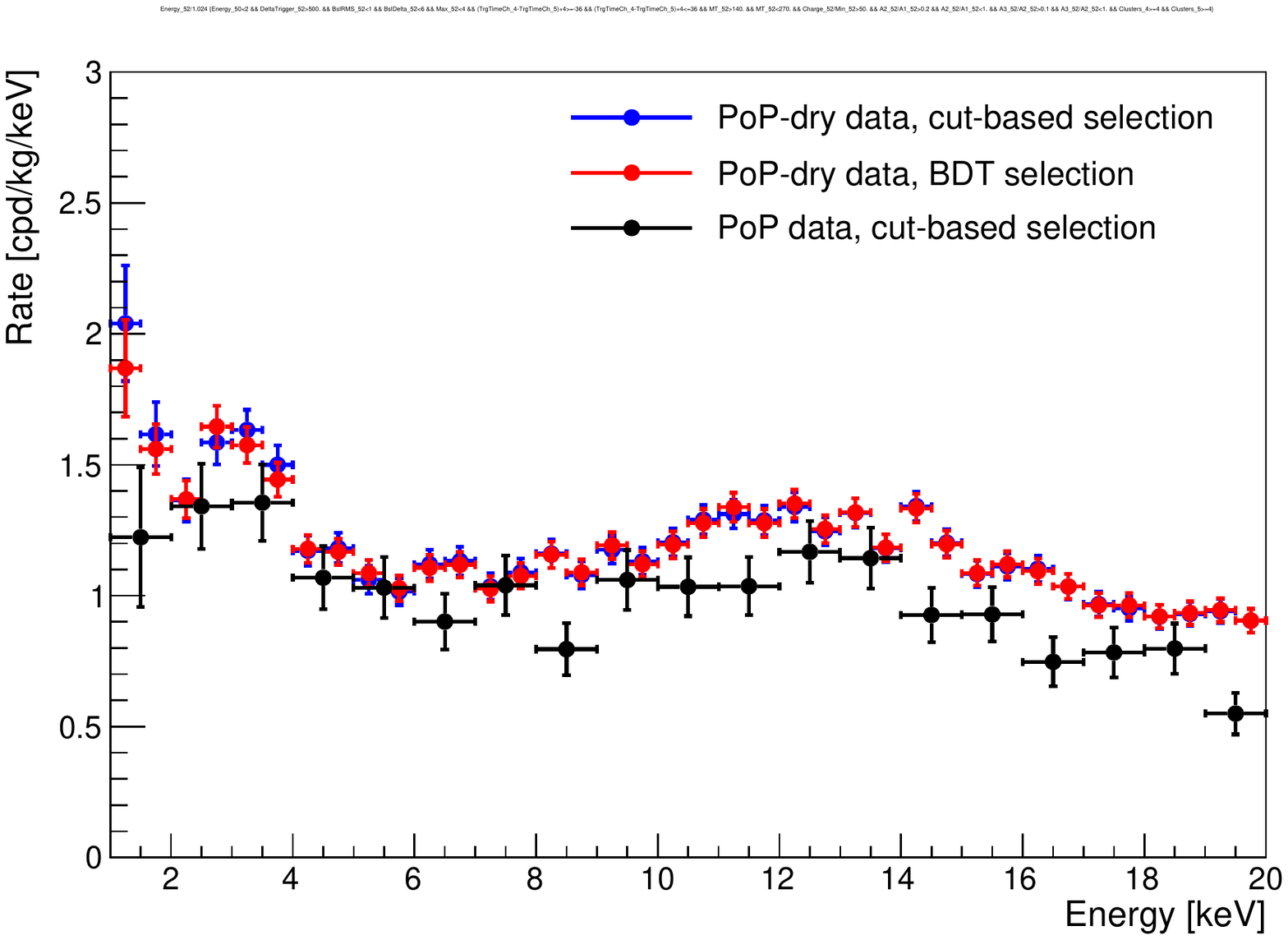}
    \caption{NaI-33 low energy spectrum (after selection cuts and acceptance-correction) for data acquired with the PoP-dry (red points for BDT and blue point for cut-based selection) and the PoP setup (black points). The latter includes only events in anti-coincidence with the liquid scintillator veto. The wider binning (and error bars) reflects the $\sim$~10 times lower exposure.}
    \label{fig:Comparison}
\end{figure}

Fig.~\ref{fig:Comparison} shows the NaI-33 energy spectrum below 20~keV acquired within the PoP-dry setup and obtained by applying the BDT (red points) or the cut-based selection (blue points). The spectrum acquired within the PoP setup (black points) and which benefits of the anti-coincidence with the liquid scintillator veto is also shown as a comparison.  
All the energy spectra are acceptance-corrected according to the applied event selection criterion.
It must be noted that the exposure collected with the PoP-dry setup is 891~kg$\cdot$days, while the exposure of PoP setup was only $\sim$~90~kg$\cdot$days. 
The resulting average count rate in the 1-6 keV ROI is 1.42 $\pm$ 0.02 cpd/kg/keV by applying the cut-based selection, and 1.39 $\pm$ 0.02 cpd/kg/keV by applying the BDT selection.
This background level is comparable to that measured within the PoP setup, i.e. 1.20 $\pm$ 0.05 cpd/kg/keV \cite{sabre-pop-2021}.
Such result demonstrates that, as the vetoable crystal internal contaminants (e.g. $^{40}$K) are low enough, the active veto is no longer necessary, for a future SABRE North detector, in order to tackle the challenge of obtaining a background rate lower than 1 cpd/kg/keV. 

Similarly to the spectral analysis performed for the PoP data, we have performed the fit of the PoP-dry energy spectrum with several background components. The spectral shapes are obtained with our Monte Carlo~\cite{sabre-mc} simulation. The activities of the following components were included: $^{40}$K, $^{210}$Pb, $^{3}$H, $^{226}$Ra, $^{232}$Th, $^{129}$I, and a flat component which accounts for $^{87}$Rb and other internal and external contributions (such as radioactivity in PMTs)\footnote{For what concerns another long lived cosmogenic isotope $^{22}$Na of some relevance for other NaI(Tl) experiments, we find no evidence in our spectrum of the high energy gamma emission and we do not include it in the fit.}.
A specific $^{238}$U contribution from the PMTs quartz window was included to better reproduce the experimental energy spectrum around 16 keV. In addition, $^{210}$Pb from the PTFE reflector wrapping the crystal was included to reproduce the peak at $\sim$~12 keV due to X-rays from $^{210}$Pb. 
All rates are allowed to vary freely with the following exceptions: $^{40}$K rate has a penalty determined by the PoP-measured value (0.07 $\pm$ 0.05 mBq/kg \cite{sabre-pop-2021}); $^{226}$Ra, $^{232}$Th have penalties to the activities determined by measuring Bi-Po sequences in \cite{nai033-udg}.

\begin{figure}[!ht]
    \centering
    \includegraphics[width=.97\columnwidth]{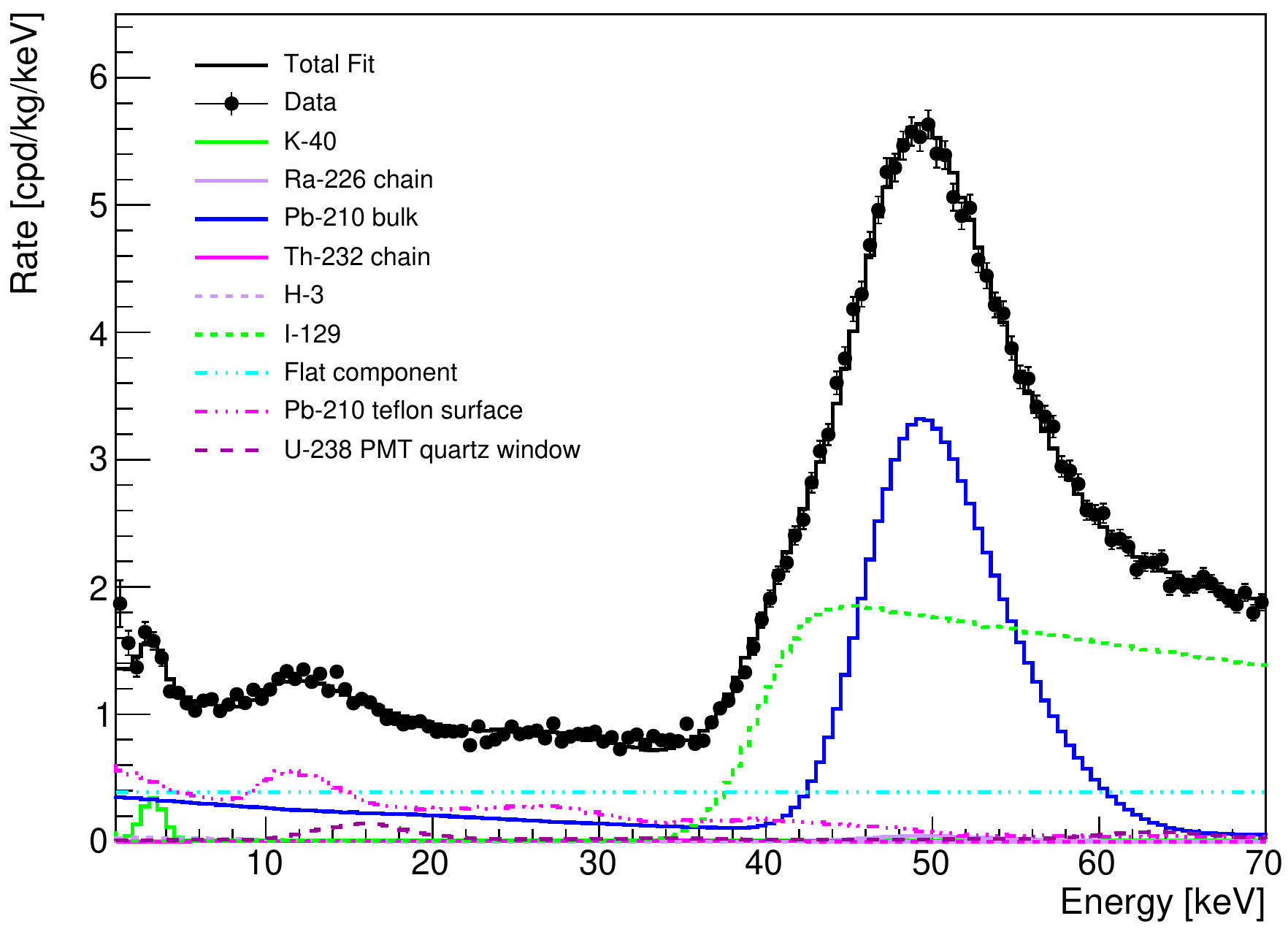}
    \includegraphics[width=.97\columnwidth]{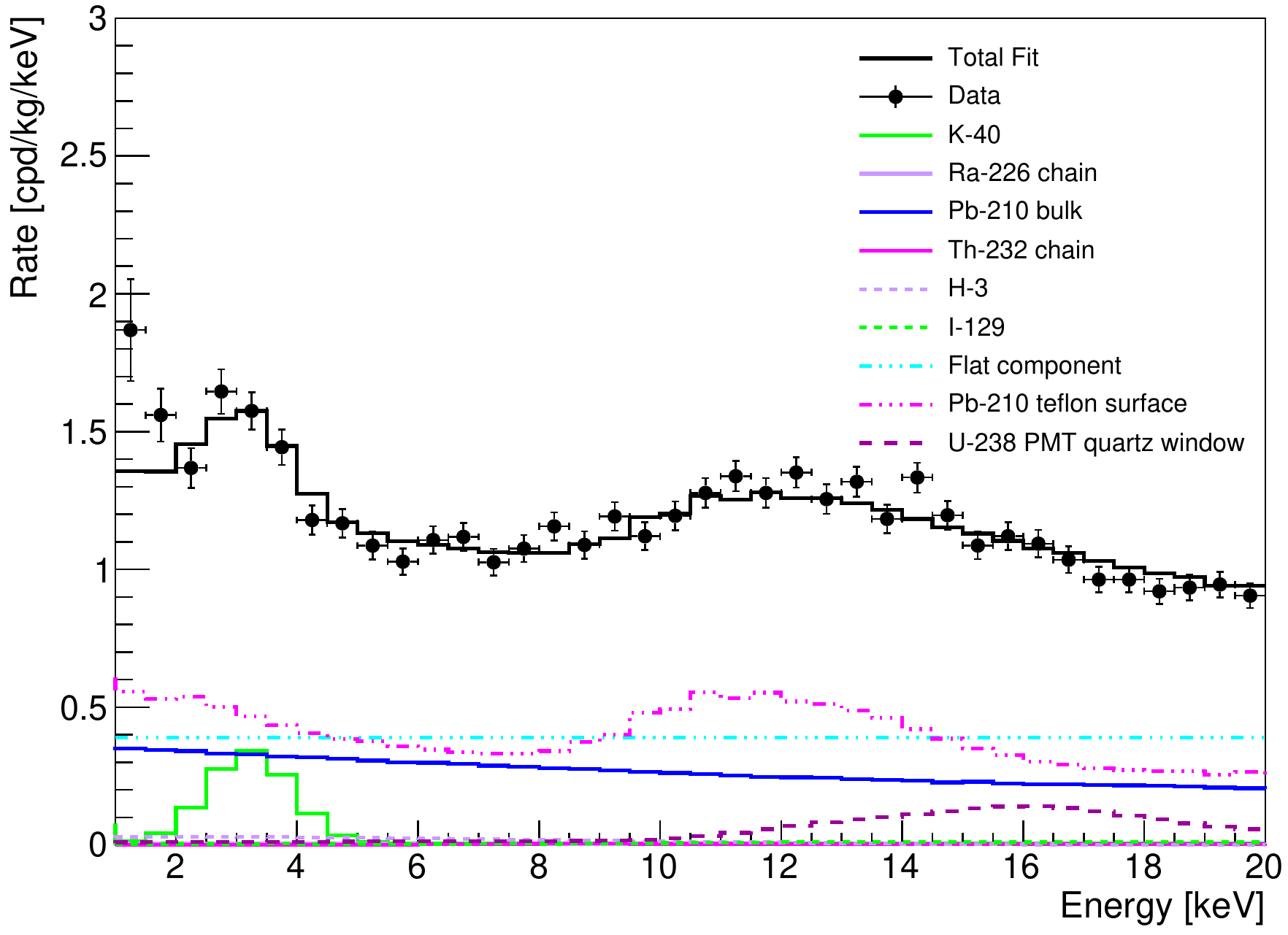}
    \caption{PoP-dry energy spectrum of the NaI-33 crystal up to 70~keV (top) and below 20~keV (bottom) with a spectral fit. Data are shown after noise rejection and acceptance correction.}
    \label{fig:NaI33SpectralFit}
\end{figure}

The result of the best-fit in the 2-70 keV energy range ($\chi^2/N_{d.o.f.} = 177/127$) is shown in Fig.~\ref{fig:NaI33SpectralFit} (extrapolating below 2~keV), while Tab.~\ref{tab:SpectralFitResults} summarises the activities of the different background components determined from the spectral fit (2$^{nd}$ column) and the rate in the 1-6 keV energy region of interest (3$^{rd}$ column). Activities from other NaI(Tl)-based experiments are also reported as a comparison (4$^{th}$, 5$^{th}$ and 6$^{th}$ column). 
The dominant background contributions in NaI-33 are from a $^{210}$Pb contamination in the PTFE reflector and from the $^{210}$Pb contamination in the crystal bulk. 
It should be noted that the PoP-dry passive shielding is not optimized for an high-sensitivity full-scale experiment. Consequently also the flat component gives a significant contribution in the ROI which we attribute to environmental gammas entering the shielding. The fit is limited to 70~keV because modeling external backgrounds above this energy would have required an effort beyond the scope of the present paper.  
A detailed Monte Carlo simulation of an optimized passive shielding geometry made of low radioactivity copper and polyethylene is under development. Preliminary results show that the background contribution in NaI(Tl) crystals from environmental gammas and neutrons, and from the radioactivity of the shielding materials is negligible with respect to that of the crystals themselves.  

\begin{table*}[!htbp]
    \centering
    \begin{tabular}{lccccc}
    \bf{Source} & \bf{Activity in} &\bf{Rate in ROI}                       &\bf{Activity in} & \bf{Activity in} 
                & \bf{Activity in} \\
                & \bf{NaI-33} & \bf{in NaI-33} & \bf {DAMA/LIBRA crystals} & \bf{ANAIS crystals} & \bf{COSINE crystal} \\
                & \bf{[mBq/kg]} & \bf{[cpd/kg/keV]} & \bf{[mBq/kg]} & \bf{[mBq/kg]} & \bf{[mBq/kg]} \\
    \hline
    $^{40}$K & 0.15$\pm$0.02 & 0.12$\pm$0.02 & $\leq$0.62 & 0.70-1.33 & 0.58-2.5 \\
    $^{210}$Pb (bulk) & 0.461$\pm$0.005 & 0.325$\pm$0.004 & 0.005-0.03 & 0.70-3.15 & 0.74-3.20 \\
    \cline{2-3}
    $^{226}$Ra & 0.0059$\pm$0.0006 & \multirow{2}{*}{0.0049$\pm$0.0005} & 0.009-0.12 & 0.003-0.01 & 0.0002-0.001 \\
    $^{232}$Th & 0.0016$\pm$0.0003 & & 0.002-0.03 & 0.0004-0.004 & 0.001-0.01 \\
    \cline{2-3}
    $^3$H & $\leq$0.005 & \multirow{2}{*}{$\leq$0.05} & $\le$0.09 & 0.09-0.20 & 0.05-0.12 \\
    $^{129}$I & 1.29$\pm$0.02 & & 0.96$\pm$0.06 & 0.96$\pm$0.06 & 0.72-1.08 \\
    \cline{2-3}
    $^{210}$Pb (PTFE) & 0.83$\pm$0.06 mBq & 0.46$\pm$0.03 & - & 0-3 mBq & 0-8 mBq \\ 
    $^{238}$U (PMTs quartz window) & 0.31$\pm$0.05 mBq & 0.011$\pm$0.002 & - & - & - \\ 
    Other (flat) & & 0.39$\pm$0.02 & & & \\
    \hline
    \bf{total} & & 1.36$\pm$0.04 & & & \\
    \hline
    \end{tabular}
    \caption{Background components in NaI-33 from the spectral fit of PoP-dry data and current rate in ROI~(1-6~keV), compared to other NaI(Tl)-based experiments \cite{dama2020-summary, dama-apparatus, anais2019-bkgmodel, cosine2018-bkgmodel, cosine2021-bkgmodel}. 
    Unlike other experiments, we did not make assumptions about the equilibrium of the $^{238}$U chain and attributed a penalty in the fit for $^{226}$Ra to the activity determined by Bi-Po measurement. The secular equilibrium is assumed instead for the $^{232}$Th decay chain by all experiments. 
    The range of values for $^{210}$Pb (PTFE) in the COSINE experiment is derived from Fig.8 of \cite{cosine2018-bkgmodel} by multiplying for the mass of the crystals.     
    Upper limits are given as one-sided 90\% CL. The total rate is conservatively calculated using upper limits.}
    \label{tab:SpectralFitResults}
\end{table*}

\section{Conclusions}

In this work, the ultra-radiopure NaI(Tl) SABRE crystal NaI-33 was characterized in the PoP-dry, a modified SABRE PoP setup that does not feature the use of a liquid scintillator veto. 
We have improved the analysis that now corrects for PMT afterpulses and performs noise rejection based on a Boosted Decision Tree with over 90\% efficiency in preserving scintillation signals.  
We find a count rate in the 1-6 keV ROI of 1.39 ± 0.02 cpd/kg/keV: a minor increase with respect to the 1.20 ± 0.05 cpd/kg/keV observed in the PoP, considering that the PoP-dry shielding leaves room for sizable improvement in the next future. 
We have obtained the background model fitting the PoP-dry energy spectrum with a combination of several Monte Carlo simulated components.
We find internal background sources consistent with the model obtained from the PoP short run.
The dominant contribution is actually not affected by a veto and can be ascribed to $^{210}$Pb: this is present in the crystal bulk, but it mostly spawns from a significant contamination in the PTFE reflector.

We are progressing toward the physics phase of SABRE North acting on several fronts. We will replace the contaminated PTFE with a radioclean one that we have selected, screened and procured.  
We are planning the growth of a crystal that benefits from the NaI powder purification with the zone refining method that we have developed \cite{sabre-pop-2021} primarily to address the bulk $^{210}$Pb. 
We are designing the detector based on an array of crystals within an improved passive shielding, compliant with the safety and environmental requirements of Laboratori Nazionali del Gran Sasso.

\section*{Acknowledgments}

This work was supported by INFN funding and National Science Foundation under the Awards No. PHY-1242625, No. PHY-1506397, and No. PHY-1620085.
We thank the Gran Sasso Laboratory for the support during the installation of the SABRE PoP setup.



 \bibliographystyle{spphys}
 \bibliography{reference}
%

\end{document}